# RESOLUTION LIMITS FOR RESONANT MEMS SENSORS BASED ON DISCRETE RELAY FEEDBACK TECHNIQUES

J. Juillard[(1)], E. Colinet[(1)(2)], M. Dominguez[(3)], J. Pons[(3)], J. Ricart[(3)]

[(1)] Département SSE, SUPELEC, France
[(2)] LETI/DCIS/SCME/LMEA, CEA, France
[(3)] MNT, UPC, Spain

## ABSTRACT

This paper is devoted to the analysis of resonant MEMS sensors based on discrete relay feedback techniques. One drawback of such techniques is that some synchronization usually occurs between the discrete part and the continuous part of the system: this results in sensor responses that are very similar to the curves known as devil's staircases, i.e. the oscillation frequency does not vary smoothly with the sensor's natural pulsation. The main contribution of this paper is a theoretical calculation of the resolution of such systems. The resolutions of two existing resonant MEMS architectures are then calculated and these results are discussed.

## 1. INTRODUCTION

In the past few years, resonant sensing has become a very popular method for measuring physical phenomena. This is especially true in the field of MEMS sensors since the electronics required for bringing a system to a state of oscillation can be fairly simple and lend themselves very well to integration. [1-4]

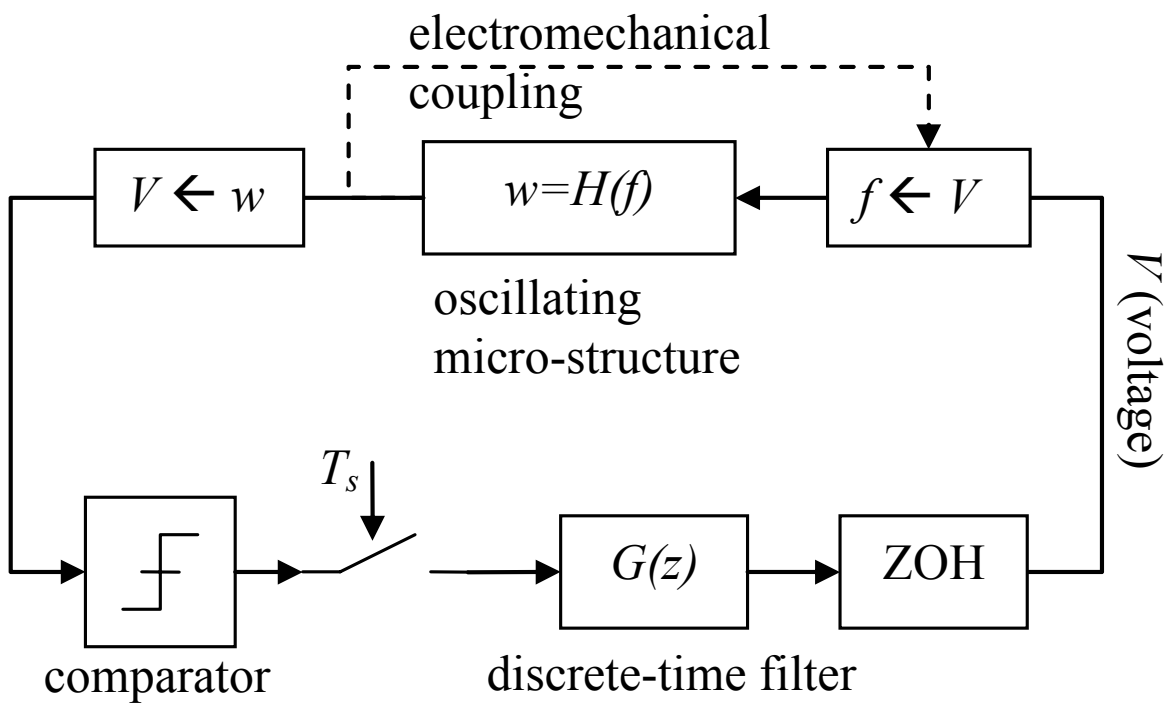

*fig. 1: block diagram of a resonant MEMS sensor using discrete relay feedback. The oscillating micro-structure can generally be modelled as a nonlinear relationship between its inputs (forces f) and its outputs (displacements, w).*

This paper is devoted to resonant MEMS sensors based on discrete relay feedback techniques: the idea is to insert the micro-structure in a feedback loop consisting of a comparator and of a discrete-time filter (fig. 1). The comparator then acts as a simple one-bit ADC (Analog-to-Digital Converter). The main advantage of this technique is that even very nonlinear position sensing techniques may be used with it, since only the sign of the position is required in order to bring the system to oscillate. Also, when the feedback filter is appropriately chosen, it is possible to greatly reduce the influence of nonlinear actuation schemes, such as electrostatic actuation [5].

One characteristic of such a technique is that some synchronization (i.e. phase-locking) usually occurs between the discrete part and the continuous part of the system: this yields the sensor responses shown in fig. 2, which are very similar to the curves known as devil's staircases [6]. Thus, such a sensor's resolution is given by the width of the steps in the sensor's response.

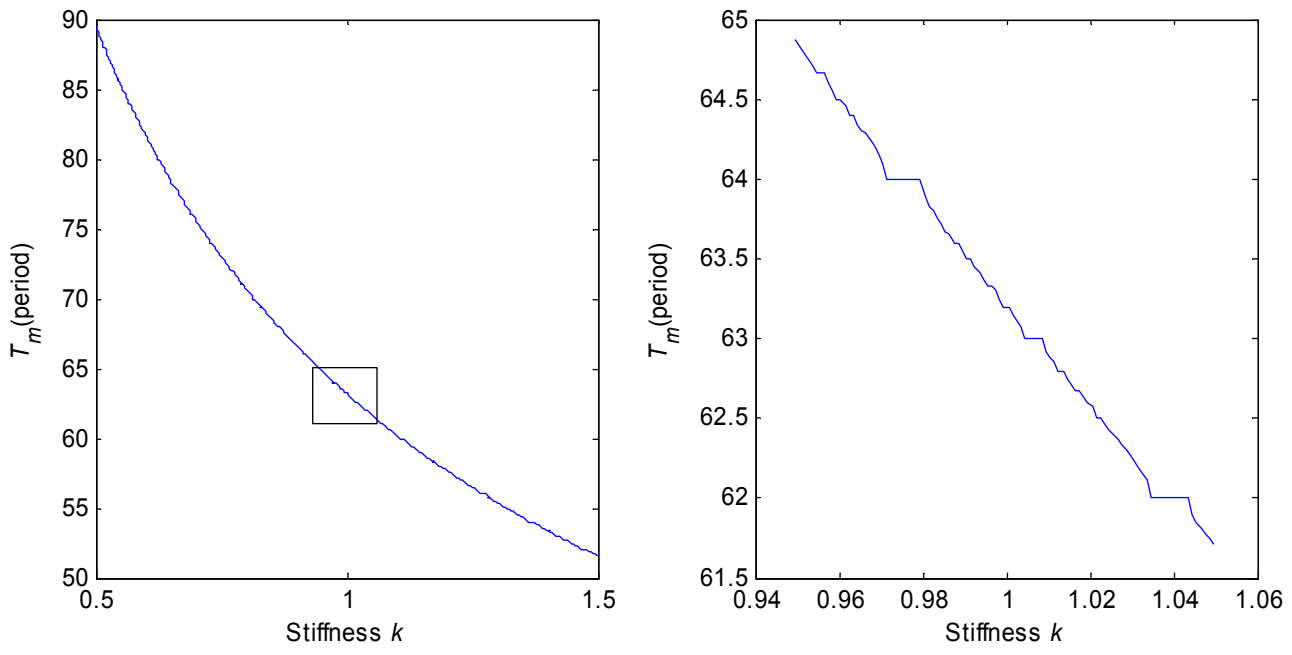

*fig. 2: typical sensor response (oscillation period vs. stiffness) for a resonant sensor using discrete relay feedback. The curve on the right is a close-up of the boxed area on the left. The sensor's resolution is limited by the size of the steps in this curve.*

These phenomena have been observed by several authors in the context of MEMS, such as Dominguez [5], Colinet [4]. They have also been studied in [6] in the somewhat different context of sigma-delta modulation. A theoretical study of these phenomena was made in [7-8], in the context of resonant MEMS sensors. In particular, the problem of determining the size of the steps in the sensor's response has been addressed in [8] in the case of linear and nonlinear systems: this is of particular importance, since these steps have a direct consequence on the sensor's resolution.

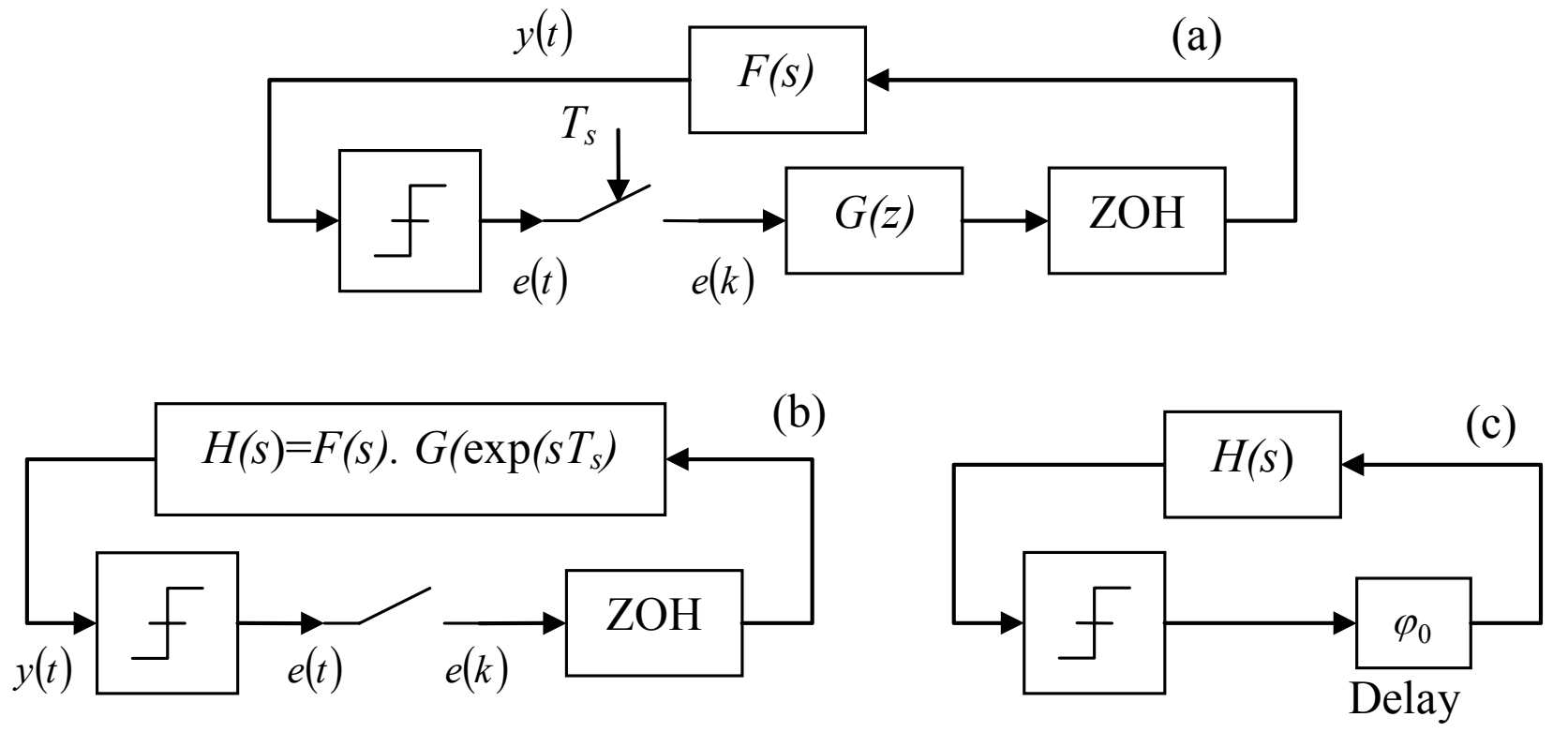

*fig. 3: the systems of (3-a) and (3-b) have the same oscillation periods $T_m$. When $T_m$ is an even integer multiple of $T_s$, the sampling-and-holding stage of (3-b) can be replaced by an equivalent delay (3-c), whose value belongs to [0, $T_s$].*

However, it is interesting to note that, depending on the considered application, these steps may become a desirable feature. This is true in the context of resonant actuation, for example of scanning micro-mirrors [9], where one tries to obtain a (large-stroke) sinusoidal displacement [5].

In the first part of the present paper, these theoretical results are restated. On this basis, the authors then go on to compare the sensitivities of the different existing actuation techniques, with a focus on those developed in [4-5], and discuss their respective advantages and drawbacks.

## 2. PROPERTIES OF MIXED-SIGNAL RELAY FEEDBACK SYSTEMS

Depending on what feedback filter *G(z)* is used, the mixed-signal system of fig. 1 may oscillate, with period $T_m$. As opposed to the case when a continuous-time feedback filter is used, a variation in the parameters of the oscillating micro-structure does not always result in a variation in the oscillation period: this corresponds to the steps appearing in fig. 2. Moreover, depending on the system's initial state, several oscillation regimes, with different values of $T_m$, may be reached. These oscillation regimes and their determination have been studied in [6-7]. An important characteristic of these systems is that the ratio of the measured period (defined as the average time between two rising edges of the comparator) to the sampling period $T_s$ is a rational number. The largest steps in the system's response occur when this ratio is an even integer.

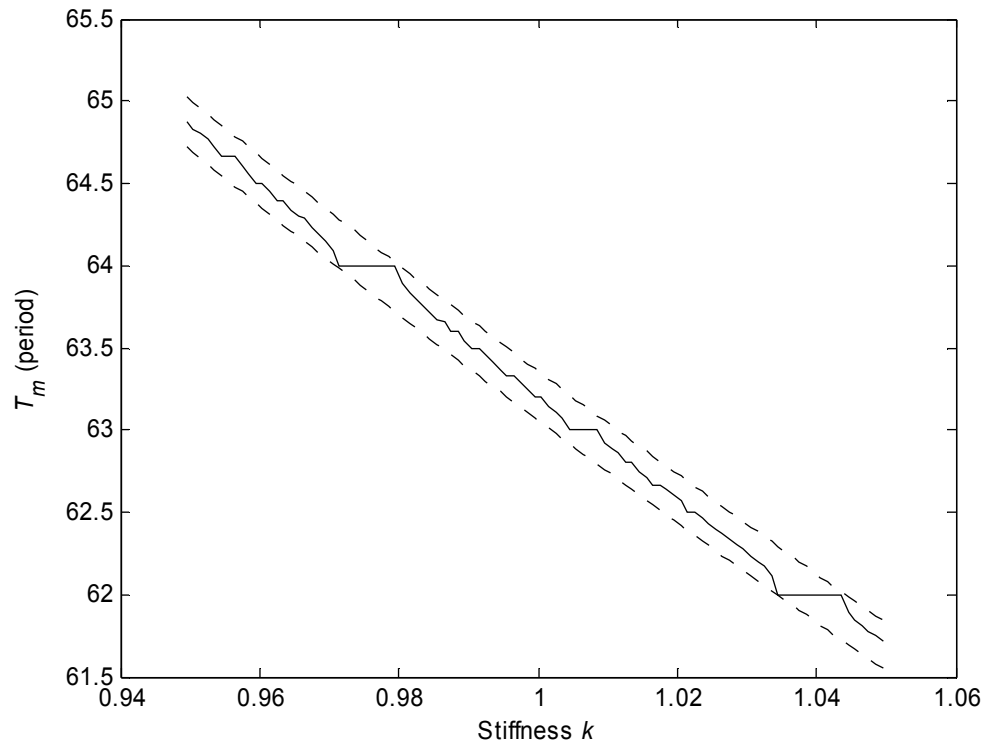

*fig. 4: the size of the steps in the sensor's response (and thus the sensor's resolution) is given by the two dashed curves. These curves can be determined [8] by calculating the period of the oscillations of the continuous system of fig. 3-c, for $\phi_0$=0 and $\phi_0$=$T_s$. The size of the smaller steps can be obtained in a similar fashion.*

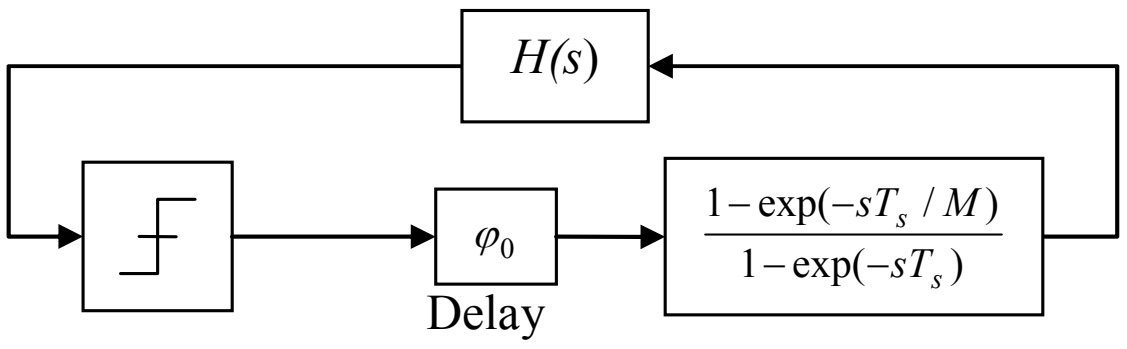

*fig. 5: when the DAC delivers pulses of length $T_s/M$, the system of fig. 3-c must be replaced by this one. The resolution of the original mixed-signal system is then obtained from the curves corresponding to $\phi_0$=0 and $\phi_0$=$T_s$.*

In [5], the authors determine the size of the steps assuming a sinusoidal wave at the comparator's input: this approximation is valid for over-sampled systems, with large quality factors. In [8], a more general manner of determining the different step-sizes is established: it is valid for most linear systems and for some nonlinear systems, provided no DC component is introduced in the loop. It is shown that, for a given set of parameters, the possible values of $T_m$ for the mixed signal system belong to the set $[T_0,T_1]$ (figs. 3 and 4), where $T_0$ and $T_1$ can be determined with simple continuous-time methods, as Tsypkin's [10] or Hamel's method [11]. It is also shown in [8] that the size of the widest steps in the sensor's response can be obtained in a straightforward fashion from the continuous-time system of fig. 3-c.

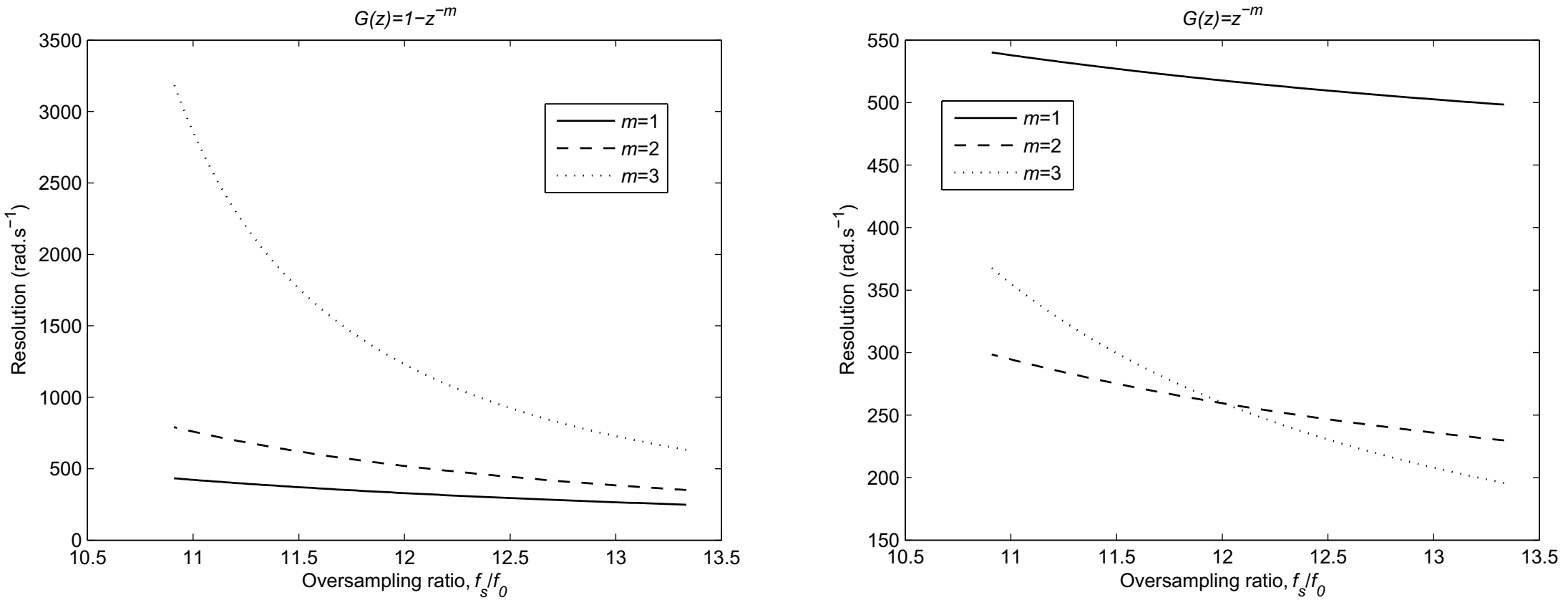

*fig. 6: as m increases, the resolution of [4] decreases (6-a) whereas that of [5] globally increases (6-b). For the considered oversampling ratios, the resolution of the PDO is usually better than that of [4].*

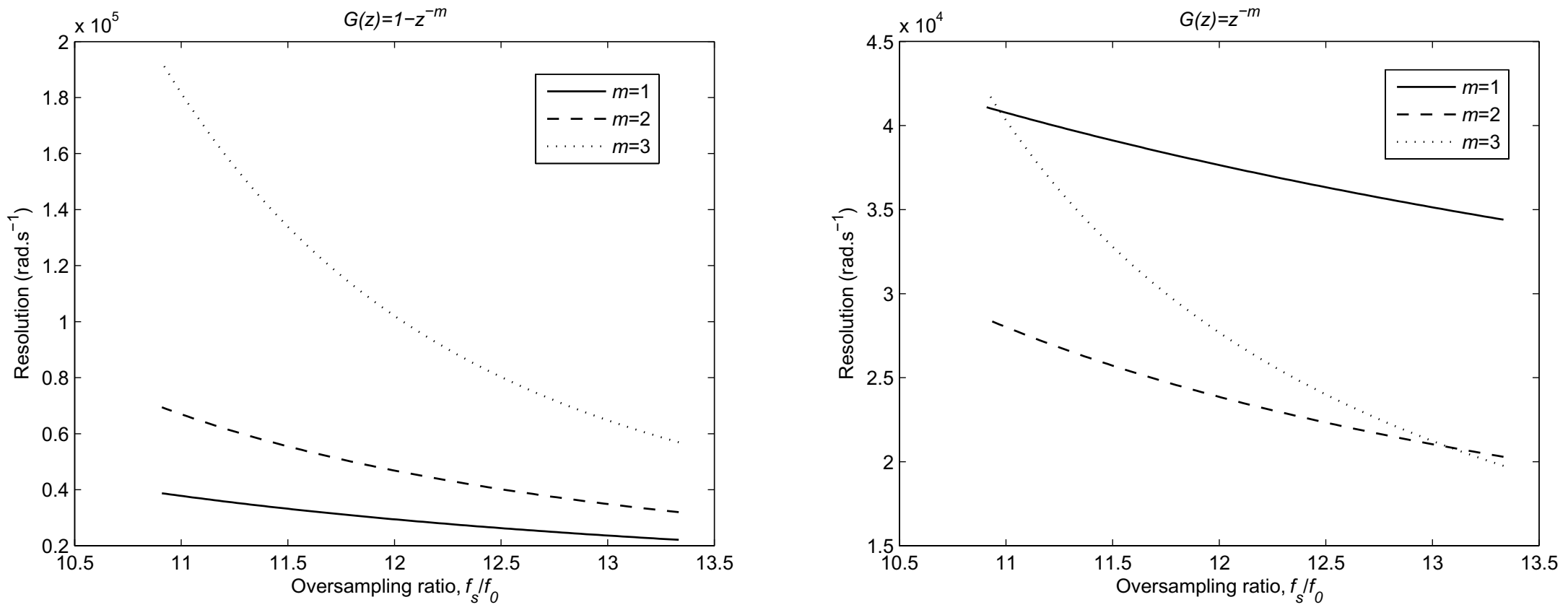

*fig. 7: as expected, a lower quality factor results in an important decrease in the sensor's resolution. For the considered oversampling ratios, the resolution of the PDO is usually better than that of [4].*

In the case the digital-to-analog converter (DAC) is not a zero-order hold but delivers pulses of width $T_s/M$, $M>1$, these results may easily be extended by using the continuous time equivalent represented in fig. 5.

## 3. COMPARISON OF TWO ACTUATION SCHEMES BASED ON MIXED RELAY FEEDBACK ARCHITECTURES

These results may be applied to the mixed-signal architectures proposed in [4] and [5]. In [4], an actuation scheme based on a discrete differentiator:

$$G(z)=1-z^{-m} \qquad (1)$$

was presented, supposing the linear part can be modelled as a second-order system:

$$F(s)=\frac{G}{\omega_0^{\,2}+2\xi\omega_0 s+s^2}\,. \qquad (2)$$

Its main advantage is that, using this scheme with an electrostatic gap-closing actuation makes it possible to obtain very large stroke displacements. More generally, there is no crosstalk between the actuation scheme and any displacement-dependent non-linearity. It is also quite simple to integrate because it only requires very low-cost components (a 1-bit ADC and a 1½ bit DAC).

In [5], an architecture based on a simple discrete delay:

$$G(z)=z^{-m} \qquad (3)$$

was presented: the PDO (Pulsed Digital Oscillator). Provided the DAC delivers short pulses of voltage, this architecture has almost the same properties as the previous one.

### 5.1. Resolution for a high quality factor

The following set of parameters is used (from [4]). The nominal natural frequency of the system is $F_0 = 35.8 \times 10^3 Hz$, the sampling frequency is equal to $F_s = 12F_0$ and the quality factor is equal to $Q = 250$.

Setting $M = 1$ and $m = 1,2 \text{ or } 3$ and letting the system's natural pulsation vary around its nominal value, we use Tsypkin's method on the equivalent continuous-time systems (fig. 3) and, using the results of [8], we obtain the results of fig. 6.

### 5.2. Resolution for a low quality factor

The parameters given in the previous sub-section correspond to a micro-structure vibrating far from the substrate. In the case of electrostatic actuation, supposing no vacuum is enforced, the quality factor should considerably decrease and the curves of fig. 6 should substantially be modified. For a quality factor $Q = 2.5$, the results of fig. 7 are obtained.

### 5.3. Influence of pulse-width on resolution

The width of the pulse delivered by the ADC should have a strong influence on the resolution of these architectures. For example, using a shorter pulse-width with the architecture of [4] is almost similar to increasing the oversampling ratio. Thus, from the tendency of the curves in figs. 6 and 7, this should result in a better resolution.

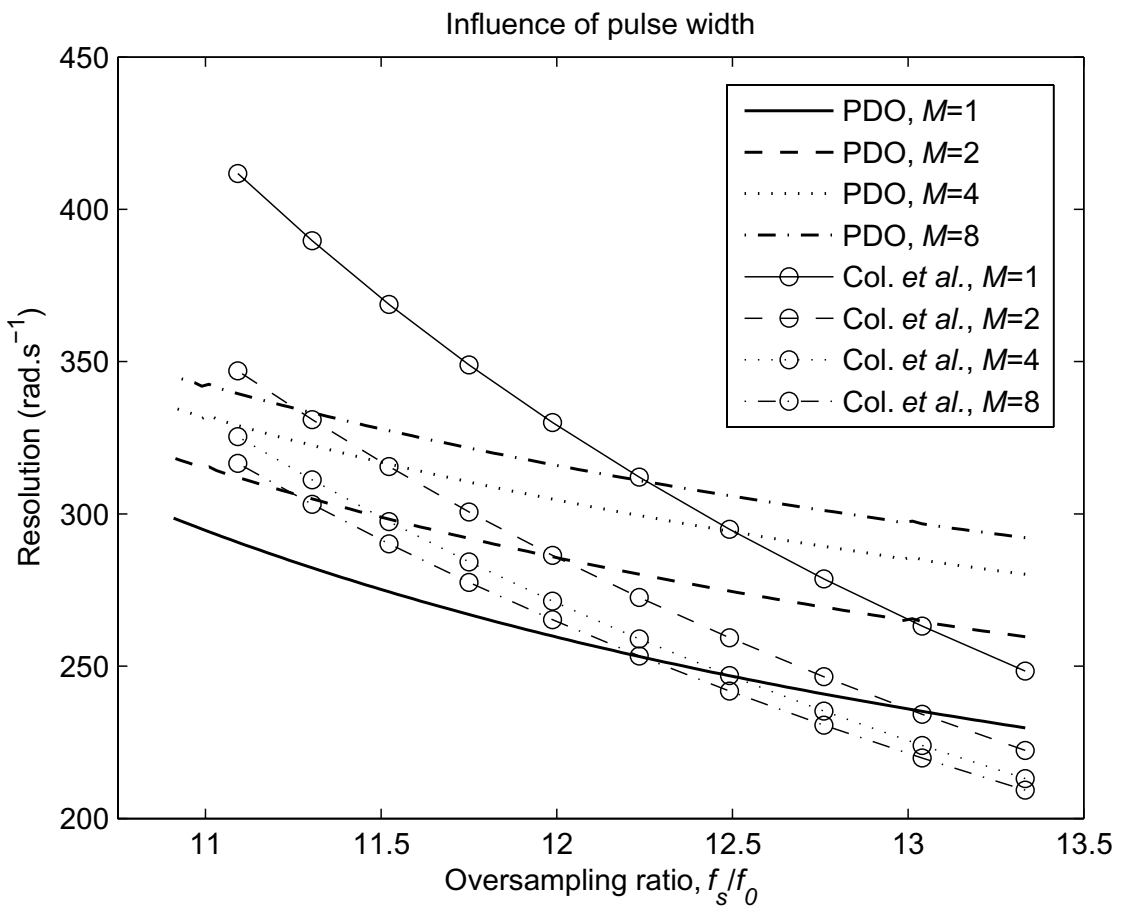

*fig. 8: the resolution of the differentiator-based scheme [4] can be improved by using shorter pulses. On the other hand, the resolution of the PDO decreases as M increases.*

On the other hand, it is more difficult to predict accurately what impact a smaller pulse-width shall have on the PDO's resolution. The results of [8] can nonetheless be applied, by using Tsypkin's method on the equivalent continuous-time systems (fig. 5).

Letting $Q = 250$, $m = 1$ for the differentiator-based scheme and $m = 2$ for the PDO, the results of fig. 8 are obtained.

## 4. RESULTS

Using some properties of mixed-signal relay feedback architectures, we have calculated and compared the maximal resolution that could be obtained with two different actuation schemes. In the context of resonant sensors, the architecture based on a discrete delay [5] seems to perform better than that based on a discrete differentiator [4]. However, it must be noted that, for pulse-widths shorter than one sampling period, the differentiator-based scheme seems to perform better than the PDO. Moreover, using the PDO with a pulse-width equal to one sampling period does not permit large-stroke actuation, as opposed to the scheme proposed in [4]. In the context of resonant actuation, where one would like to obtain a very stable oscillation frequency, i.e. a very low sensitivity to variations in the micro-structure's parameters, the differentiator-based scheme seems to perform quite well, as does a PDO with short pulses.